\documentclass[pre,10pt, twocolumn]{revtex4-1}

\usepackage{graphicx}
\usepackage{amsmath}
\usepackage[utf8]{inputenc}
\usepackage{hyperref}
\usepackage{url}
\usepackage{amssymb}
\usepackage{amsfonts}
\usepackage{indentfirst}
\usepackage[english]{babel}
\usepackage[toc,title]{appendix}
\usepackage{color}
\usepackage{bm}
\newcommand{\zb}{\bar{z}}

\begin{document}

\title{Real spectra of large real asymmetric random matrices}
\author{ Wojciech Tarnowski}
\email{wojciech.tarnowski@doctoral.uj.edu.pl}
\affiliation{Institute of Theoretical Physics, Jagiellonian University, 30-348 Cracow, Poland}
\date{\today}

\begin{abstract}

When a randomness is introduced at the level of real matrix elements, depending on its particular realization, a pair of eigenvalues can appear as real or form a complex conjugate pair. We show that in the limit of large matrix size the density of such real eigenvalues is proportional to the square root of the asymptotic density of complex eigenvalues continuated to the real line. This relation allows one to calculate the real densities up to a normalization constant, which is then applied to various examples, including heavy-tailed ensembles and adjacency matrices of sparse random regular graphs.
\end{abstract}
\maketitle

\section{Introduction}  
It is almost impossible to imagine a quantitative field, in which a diagonalization of non-symmetric matrices is not used.
Following the pioneering works of Wigner~\cite{Wigner} and Dyson~\cite{Dyson}, spectra of random matrices have been intensively studied in physics and mathematics. Ginibre initiated studies of random matrices lacking symmetries~\cite{Ginibre}, but real matrices turned out to be the most difficult path on the Dyson threefold way~\cite{Dyson3w}, and it took more than 40 years to pave the way for fully solving the real Ginibre ensemble~\cite{LehmannSommers,Edelman1,Edelman2,AkemannKanzieper,Sinclair,BorodinSinclair,ForresterNagano,BorodinSinclair2,SommersWieczorek}. 

\begin{figure}
\includegraphics[width=0.48\textwidth]{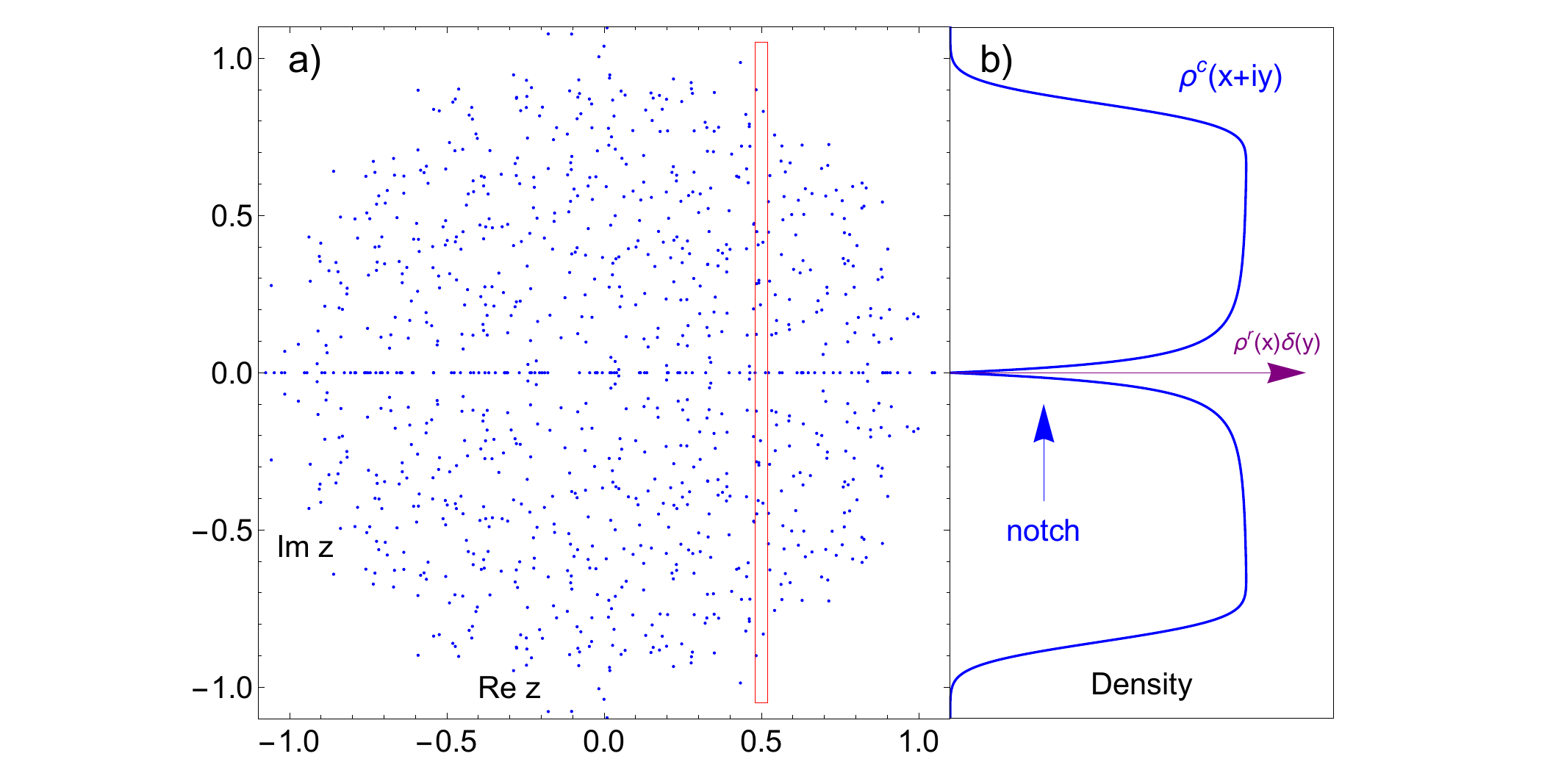}
\caption{(a) Eigenvalues of 10 random real Ginibre matrices of size $N=100$. (b) Section of density of complex eigenvalues at $x=0.5$, corresponding to the histogram obtained by collecting large number of eigenvalues falling into the red rectangle. Spike at $y=0$ is a consequence of accumulation of real eigenvalues, while the notch in the complex density originates from the repulsion between real and complex eigenvalues.
\label{Fig:1}
}
\end{figure}

Density of real asymmetric random matrices consists of two components: density of purely real eigenvalues $\rho^r(x)$ and the density of complex eigenvalues $\rho^c(z)$. Since the number of real eigenvalues of random matrices grows slower than the matrix size~\cite{Edelman1}, $\rho^r(x)$ is subleading in the large $N$ limit. Nevertheless, for matrix sizes encountered in practice, real eigenvalues significantly mark their presence in the spectrum by condensing on the real axis and repelling other eigenvalues from its vicinity, see Fig.~\ref{Fig:1}.

Although there exist many analytic techniques for evaluating $\rho^c(z)$ in large $N$, including Feynman diagrams~\cite{Janik1,FeinbergZee,ChalkerWang}, free probability~\cite{Janik2,BelinschiSpeicher,MingoSpeicher} and cavity equations~\cite{NeriMetz,MetzNeriRogers}, there is no systematic way of calculating the asymptotic density of real eigenvalues if the asymptotic density of complex eigenvalues touches the real line, as it is in the Ginibre ensemble (see Fig.~\ref{Fig:1}). All known  formulas for $\rho^r(x)$ are obtained by laboriously taking the $N\to\infty$ limit in exact expressions for finite size~\cite{Edelman1,ForNagaoElliptic,TruncOrt,Simm1,Simm2}, but this procedure is limited to exactly solvable ensembles.

This notorious difficulty can be explained with the two arguments. 
First, a standard way of calculating the density relies on embedding the support of eigenvalues into a higher-dimensional space (complex for real spectra and quaternions for complex spectra~\cite{NowakJarosz}), evaluate the Green's function in the extended space and study its behavior in the vicinity of the spectrum. Density of real eigenvalues should then be embedded into a complex space, but the complex-valued Green's function is then not defined in the vicinity of the real line, due to the presence of complex eigenvalues nearby. If the support of the asymptotic density of complex eigenvalues is separated from the real line, these standard tools can be applied, see e.g.~\cite{FeinbergRiser}. Second, the density of real eigenvalues is a $1/\sqrt{N}$ correction to the asymptotic density~\cite{Edelman1}, thus \textit{a priori} inaccessible by perturbative $1/N$ expansion using Feynman diagrams.

In this work, assuming additionally orthogonal invariance of the probability of matrix elements, we show that if the density of complex eigenvalues extends to the real axis, i.e., real eigenvalues can switch into complex ones, depending on the realization of randomness, 
the asymptotic densities of real and complex eigenvalues are related via a remarkably simple formula 
\begin{equation}
\rho^r(x) \sim \sqrt{\rho^c(z=x+0i)}. \label{eq:Main}
\end{equation}

\section{Real partial Schur decomposition} 
Let $x_1$ and $x_2$ be real eigenvalues of a $N\times N$ real matrix $X$. Then, it can be represented as $X=O_1 X_1 O_1^T$, where 
\begin{equation}
\sbox0{$\begin{matrix}x_1 & t_{12}
\\ 
0 & x_{2}\end{matrix}$}
X_1 = \left(
\begin{array}{c|c}
\usebox{0}&\makebox[\wd0]{\large $T_1$}\\
\hline
  \vphantom{\usebox{0}}\makebox[\wd0]{\large $0$}&\makebox[\wd0]{\large $Y_1$}
\end{array}
\right)
\label{eq:SchurReal}
\end{equation}
and the matrix $Y_1$ is of size $(N-2)\times (N-2)$. The orthogonal matrix $O_1$ is obtained by a composition of two Householder reflections and the matrix $T_1$ is of size $2\times(N-2)$.

If $X$ has a pair of complex conjugate eigenvalues $z=x+iy$ and $\zb=x-iy$, the upper-left block cannot be brought to the upper-triangular form without leaving the field of real numbers. Instead, the decomposition $X=O_2 X_2 O_2^{T}$ uses slightly different form of $X_2$:
\begin{equation}
\sbox0{$\begin{matrix}x & b \\-c &x \end{matrix}$}
X_2 = \left(
\begin{array}{c|c}
\usebox{0}&\makebox[\wd0]{\large $T_2$}\\
\hline
  \vphantom{\usebox{0}}\makebox[\wd0]{\large $0$}&\makebox[\wd0]{\large $Y_2$}
\end{array}
\right). \label{eq:SchurComplex}
\end{equation}
Eigenvalues of the upper-left block are $z$ and $\zb$, thus $y^2=bc$. There remains one additional degree of freedom, namely $\eta=b-c$ is not fixed. This is a counterpart of $t_{12}$ from \eqref{eq:SchurReal}. These decompositions are the real partial Schur decompositions -- the building blocks of the iterative construction of the Schur decomposition algorithm.

If the matrix $X$ is random, its elements are given by some probability measure $P(X) dX$. The flat measure over the matrix elements transforms under the nonlinear change of variables induced by the partial Schur decomposition. In case of two real eigenvalues it reads~\cite{Edelman1}
\begin{multline}
dX = \\
|(x_1-x_2) \det(x_1-Y_1) \det(x_2-Y_1)| dx_1 dx_2 dt_{12} dY_1 dT_1 dO_1, \label{eq:Jac1}
\end{multline}
where $dO_1$ is the measure on the orthogonal matrix (see~\cite{Edelman1} for details). In case of a pair of complex conjugate eigenvalues, the flat measure transforms to~\cite{Edelman2}
\begin{equation}
dX = \frac{2|\eta y|}{\sqrt{\eta^2+4y^2}} |\det(z-Y)|^2 dx dy d\eta dT_2 dO_2 dY_2. \label{eq:Jac2}
\end{equation}
After integrating all variables but $x_1$ and $x_2$, we obtain the 2-point density of real eigenvalues $\rho^r(x_1,x_2)$, though not normalized to unity. Integrating out all variables except $x$ and $y$ in the second case, we obtain the one-point density of complex eigenvalues $\rho^c(z)$, normalized to the average number of complex eigenvalues (see also~\cite{Edelman1,Edelman2}). Orthogonal matrices $O_1$ and $O_2$ are not related to each other, but assuming additionally orthogonal invariance of the probability measure, i.e., $P(X) = P(OXO^T)$,
integration over them yields only constants.

The two-point density can be represented as a product of one-point densities and a connected density, $\rho^r(x_1,x_2) = \rho^r(x_1)\rho^r(x_2) + \rho_{conn}^r(x_1,x_2)$. At this point we use the large matrix size regime, at which we are working. Correlations between eigenvalues decay on scales much larger than the typical separation of eigenvalues ($\sim 1/\sqrt{N}$), thus the connected part of the two-point density is a subleading correction in the matrix size. This fact was shown in many models using diagrammatic methods and loop equations~\cite{AJM,BrezinZee}. Furthermore,~\cite{ForresterNagano} provides an explicit calculation of the connected density of real eigenvalues for the real Ginibre ensemble, showing its exponential decay. This result is expected to be universal in the bulk of the spectrum.  Therefore, the repulsion term $|x_1-x_2|$ in~\eqref{eq:Jac1} does not play a role in the large $N$ limit, only the product of two determinants and the initial probability density function are relevant. In the large $N$ limit the 2-point density factorizes $\rho^r(x_1,x_2) = \rho^r(x_1) \rho^r(x_2)$.

The term $\frac{2|\eta y|}{\sqrt{\eta^2+4y^2}}$ in \eqref{eq:Jac2} originates from the repulsion between a pair of complex conjugate eigenvalues as it decays to 0 when the eigenvalues come closer.  It is responsible for a notch in the (one-point) density of complex eigenvalues at the vicinity of the real line (see Fig.~\ref{Fig:1}). The size of the notch decreases with the matrix size, showing that also in this case the eigenvalue repulsion is immaterial in the large $N$ limit. Again, only the squared modulus of the characteristic polynomial and the probability density $P(X)$ are relevant. 

Furthermore, decompositions \eqref{eq:SchurReal} and \eqref{eq:SchurComplex} are almost identical. 
Although in general $P(X)$ is transformed into different forms under the real and complex partial Schur decompositions, the resulting difference stems from the upper-left $2\times 2$ block, which is expected not to play a role in the large $N$ limit, given its finite size. 
The crucial difference between the real and complex partial Schur decompositions lies in their Jacobians, cf. \eqref{eq:Jac1} and \eqref{eq:Jac2}.
Lastly, let us notice that when setting $x_1=x_2=x$ in \eqref{eq:Jac1} and $z=\zb=x$ in \eqref{eq:Jac2}, the characteristic polynomials are the same. Taking the above considerations into account, we conclude that upon taking the large $N$ limit the continuation of the density of complex eigenvalues to the real axis is proportional to the square of the density of real eigenvalues, hence the relation~\eqref{eq:Main}. 

By comparing two Schur decompositions \eqref{eq:SchurReal} and \eqref{eq:SchurComplex}, we implicitly assumed that the same pair of eigenvalues can occur as real and complex conjugate. In other words, different realizations of randomness in $X$ may lead to a pair of complex conjugate eigenvalues as well as a pair of real eigenvalues, though such events not necessarily need to be equiprobable. This scenario is not realized in ensembles with topological constraints put on eigenvalues. For example, the Perron-Frobenius eigenvalue is guaranteed to be real, thus the branch \eqref{eq:SchurComplex} never applies there. Additionally, pseudosymmetric matrices form another class of matrices violating this assumption. Let $A$ and $B$ be real symmetric matrices with $B$ positive definite. While their product $AB$ is not symmetric, its spectrum is real, because the product is isospectral with a symmetric matrix $B^{1/2}AB^{1/2}$.

Although the orthogonal invariance of the pdf of matrix elements was assumed throughout the derivation, it is not clear whether this assumption is necessary. In the next section we provide an example of an ensemble with non-invariant pdf, to which the formula \eqref{eq:Main} still applies upon taking into account topological eigenvalues. Last but not least, we remark that taking large $N$ limit is essential in the derivation, thus we do not expect the main result to hold for ensembles the density of which is obtained in a double scaling limit as e.g. in weak asymmetry regime of the elliptic ensemble~\cite{ForNagaoElliptic,FyodKhor}.


\section{Applications} 
The relation \eqref{eq:Main} determines the density of real eigenvalues up to the normalization constant. It cannot predict the expected number of real eigenvalues, thus in all considered examples we assume normalization $\int_{-\infty}^{\infty}\rho^r(x)dx=1$.

(0) All known results on the asymptotic density of real eigenvalues, which were calculated by taking the large $N$ asymptotics of exact densities, can be recovered from the knowledge of the asymptotic density of complex eigenvalues. This includes the ratio of two Ginibre matrices~\cite{Edelman1}, products of real Ginibre matrices~\cite{Simm1}, truncated orthogonal matrices~\cite{TruncOrt} and their products~\cite{Simm2,ForresterIpsenKumar}. Moreover, this relation applies also to the spectra of random Lindblad operators~\cite{Lindblads}, where it was numerically observed for the first time and inspired this study. Furthermore, it allows us to obtain a series of novel results for models in which the density of real eigenvalues was not even attempted before.

\begin{figure*}
\includegraphics[width=0.32\textwidth]{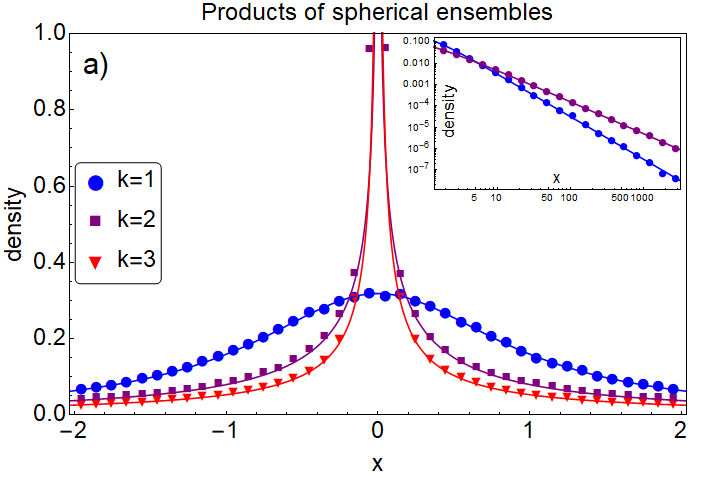}
\includegraphics[width=0.32\textwidth]{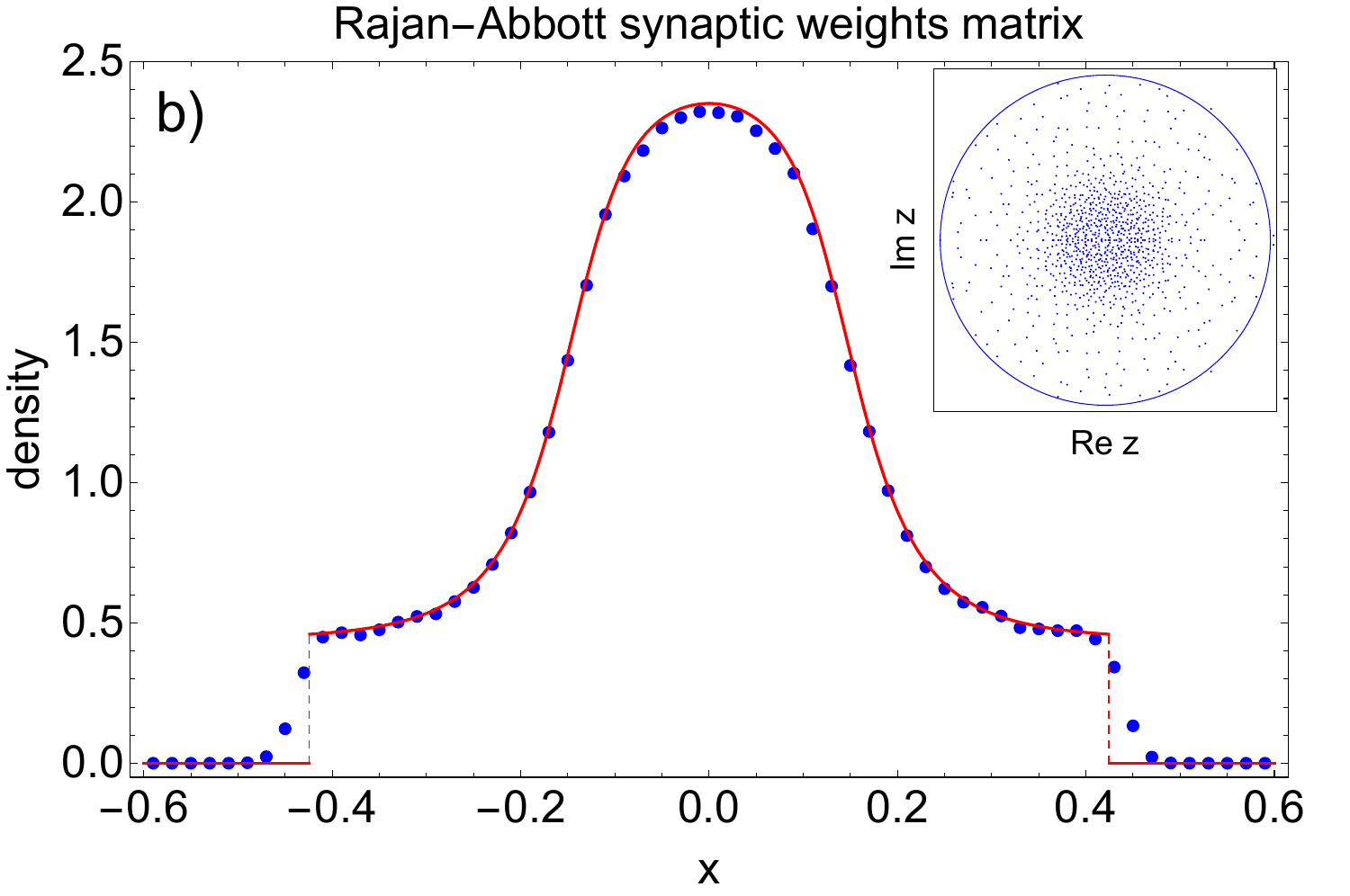}
\includegraphics[width=0.32\textwidth]{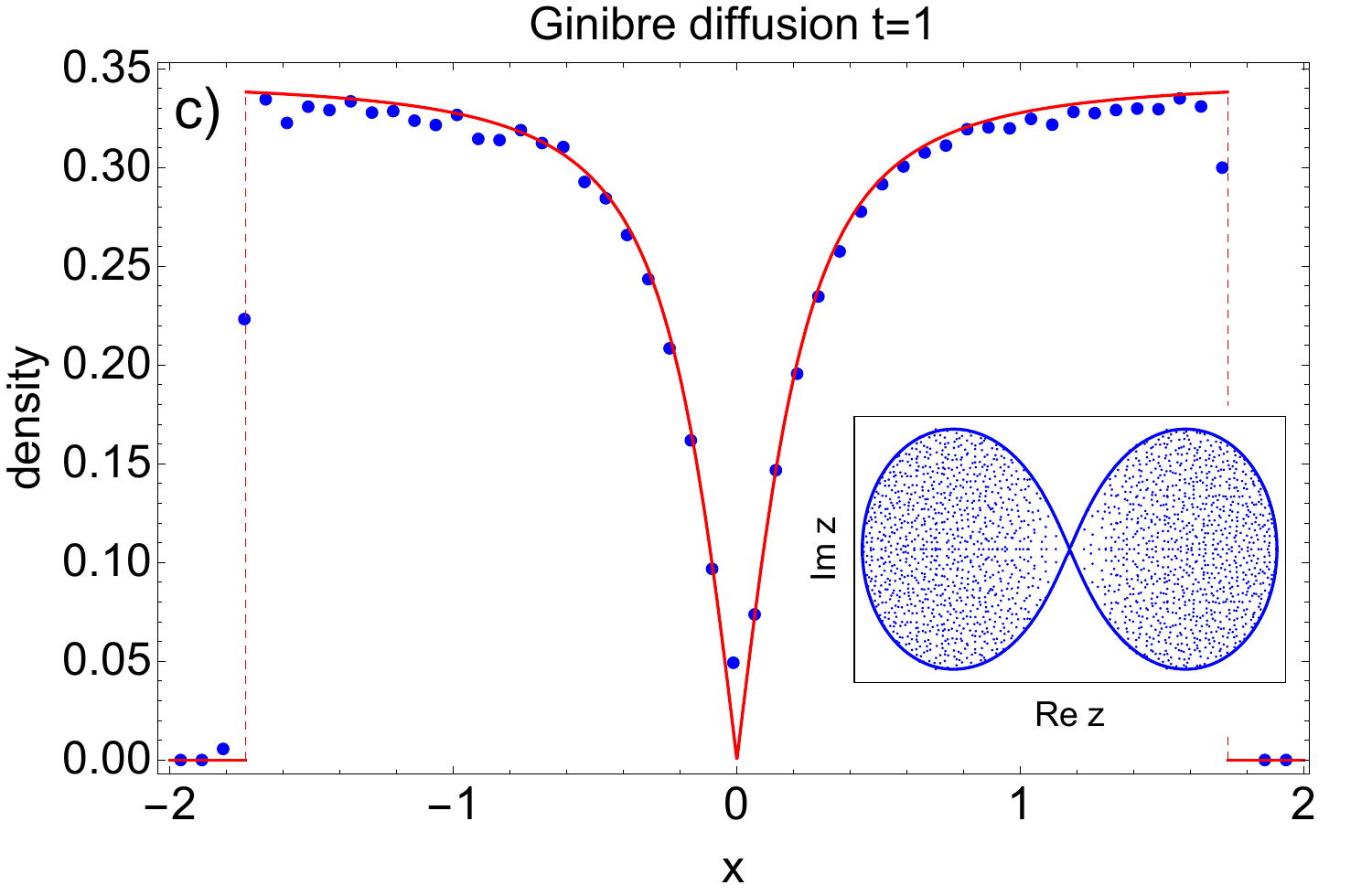}
\caption{(a) Density of real eigenvalues of products of $k$ spherical ensembles. The inset presents densities on double-logarithmic scales to emphasize power-law tails. The solid lines are given by \eqref{eq:densSph}. (b) Density of real eigenvalues of a synaptic connectivity matrix introduced by Rajan and Abbott~\cite{RajanAbbott} with parameters $\sigma_E=0.15$, $\sigma_I=0.9$ and $f_E = 0.8$. The inset presents complex eigenvalues of a single realization, bounded by the circle of radius $r^2=(1-f_E)\sigma_I^2+f_E \sigma_E^2$. (c) Density of real eigenvalues of a matrix following the real Ginibre diffusion, starting at the initial condition with half of its eigenvalues equal to 1, and half equal to -1. The solid line is given by \eqref{eq:densSpiric} with the constant $c(1)=0.6105$. The inset shows all eigenvalues on the complex plane for a single realization. All numerical results are obtained by diagonalization of $10^4$ matrices of size $N=1000$.
\label{Fig:2}
}
\end{figure*}


\begin{figure}
\includegraphics[width=0.49\textwidth]{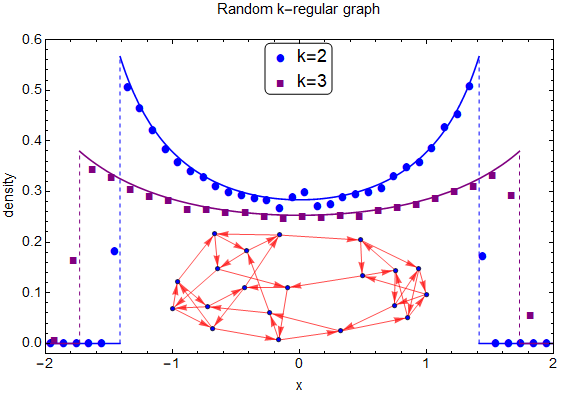}
\caption{
Density of real eigenvalues of adjacency matrices of random $k$-regular oriented graph. Numerical results obtained by diagonalization of 5000 matrices of size $N=1000$ for $k=2$ and $14000$ matrices of size $N=200$ for $k=3$. Solid lines are given by \eqref{eq:densReg}, which does not take into account the Perron-Frobenius eigenvalue $\lambda_{isol}=k$ and zero eigenvalues for 2-regular graphs. These eigenvalues were excluded from the histograms so as not to affect the normalization. The inset shows an example of a 2-regular oriented graph with 20 vertices.
\label{Fig:3}
}
\end{figure}

(1) Products of spherical ensembles. Let $X_1$ and $X_2$ be two independent Ginibre matrices. The product $Y=X_1 X_2^{-1}$ is distributed according to the spherical ensemble~\cite{Spherical}. The density of complex eigenvalues of a product of $k$ such matrices is heavy-tailed and given by~\cite{HL1} 
\begin{equation}
\rho^c(z) = \frac{1}{\pi k} \frac{|z|^{\frac{2}{k}-2}}{(1+|z|^{2/k})^2}.
\end{equation}
Therefore, the density of real eigenvalues reads
\begin{equation}
\rho^r(x) = \frac{1}{\pi k} \frac{ |x|^{\frac{1}{k}-1}}{1+|x|^{2/k}}. \label{eq:densSph}
\end{equation}
See Fig.~\ref{Fig:2} for the numerical verification.


(2) Rajan-Abbott model of synaptic connectivity matrices~\cite{RajanAbbott}. Matrix elements of $W$ are Gaussian random numbers with mean $\mu_i$ and variance $\sigma_i/\sqrt{N}$, where $i\in \{I,E\}$ and I denotes inhibitory neurons, while E stands for excitatory neurons. There are $Nf_E$ excitatory neurons and $N(1-f_E)$ inhibitory ones. One also imposes the excitatory/inhibitory balance by demanding that $f_I\mu_I+f_E\mu_E = 0$ and $\sum_{k=1}^{N} W_{jk}=0$. Then, the spectrum is insensitive to the means $\mu_i$ and the density of complex eigenvalues possesses rotational symmetry on the complex plane. The radial cumulative distribution $F(r) =2\pi \int_0^{r}\rho^c(s=|z|)sds$ can be calculated using Feynman diagrams~\cite{Wei} or free probability~\cite{Neco}. It satisfies the equation
\begin{equation}
1=\sum_{i} \frac{f_i\sigma_i^2}{r^2-\sigma_i^2 (F(r)-1)}, \label{eq:RA}
\end{equation}
which upon solving provides the density via $\rho^c(r) = \frac{1}{2\pi r} \frac{dF(r)}{dr}$. The density of real eigenvalues, which has not been studied in this model, can be immediately obtained from \eqref{eq:RA}. The normalization constant needs to be calculated numerically.

(3) Ginibre diffusion~\cite{DysonianPRL}. Let matrix elements of $X$ undergo diffusion with the diffusion constant $1/N$. Such a process, although trivial in the space of elements, induces nontrivial dynamics in the space of eigenvalues and eigenvectors~\cite{Dubach,GrelaWarchol}. Such a process has been studied only for complex matrices~\cite{DysonianPRL,Dubach,GrelaWarchol,NPB,Hall}. If as an initial condition a matrix with the spectral density $\frac{1}{2}\delta(z-1) + \frac{1}{2}\delta(z+1)$ is chosen, the spectrum is bounded by the spiric section $t(1+|z|^2) = |1-z^2|^2$ and the density reads~\cite{NPB}:
\begin{equation}
\rho^c(z) = -\frac{1}{8 \pi x^2} + \frac{1}{\pi t} + \frac{t}{8 \pi x^2 \sqrt{16 x^2 + t^2}}.
\end{equation}
The density is constant in the imaginary direction. In the real-valued diffusion the spectral density of complex eigenvalues remains the same, but there is an accumulation of real eigenvalues with the density
\begin{equation}
\rho^r(x) = c(t) \left[-\frac{1}{8 \pi x^2} + \frac{1}{\pi t} + \frac{t}{8 \pi x^2 \sqrt{16 x^2 + t^2}}\right]^{1/2}, \label{eq:densSpiric}
\end{equation}
where $c(t)$ is a normalization constant calculated numerically. 

(4) Adjacency matrices of random regular graphs~\cite{NeriMetz}. Let $G$ be a directed graph with $N$ nodes such that each node has exactly $k$ outgoing edges and exactly $k$ incoming edges and $X$ be its adjacency matrix. $G$ is sampled uniformly from the space of all graphs fulfilling this condition. Unlike in previous examples, here the source of randomness in $X$ is the topology of the graph, not the distribution of weights. The pdf of matrix elements is neither orthogonally invariant nor continuous. Spectra of adjacency matrices random graphs possess a rich structure. The density consists of an absolutely continuous part and point masses. Zero eigenvalues are a consequence of the lack of strong connectivity, while an outlier is the Perron-Frobenius eigenvalue~\cite{MetzNeriRogers,NeriLinear}.
 The spectral density of complex eigenvalues of $X$ reads~\cite{NeriMetz}
\begin{equation}
\rho^c(z) = \frac{k-1}{\pi} \left(\frac{k}{k^2-|z|^2}\right)^2,
\end{equation}
hence the density of real eigenvalues is given by
\begin{equation}
\rho^r(x) = \frac{1}{\log\frac{(\sqrt{k}+1)^2}{k-1}} \frac{k}{k^2-x^2}. \label{eq:densReg}
\end{equation}

Formula \eqref{eq:Main} does not apply to eigenvalues that are always real irrespective of the realization of the randomness. One therefore needs to discard zero modes and Perron-Frobenius eigenvalues from consideration, which also changes the normalization constant. Once these eigenvalues are dropped from the sample, \eqref{eq:Main} describes the continuous part of the density of real eigenvalues as numerically verified in Fig.~\ref{Fig:3}.

\section{Conclusions} 

We presented a remarkably simple relation linking the asymptotic density of real eigenvalues of random real asymmetric matrices with the density of their complex eigenvalues. This relation applies to eigenvalues that, depending on the realization of randomness, can be real or occur in complex conjugate pairs. While the orthogonal invariance of pdf was assumed in the derivation, there are evidences for relaxing this assumption. Proving the main result with full mathematical rigor remains an open problem. It is tempting to speculate whether it can be extended to the non-orthogonal eigenvectors, where the partial Schur decomposition opened a new direction of research~\cite{Dubach,Fyodorov,FyodorovTarnowski}

\begin{acknowledgments}
The author is grateful to S. Denisov, K. Życzkowski, D. Chruściński, G. Akemann, M. Kieburg, M. A. Nowak, I. Neri and Y. Fyodorov for discussions.
\end{acknowledgments}

\end{document}